%% file: LFpaper.tex
\documentclass{emulateapj}

\def\gs{\mathrel{\raise0.35ex\hbox{$\scriptstyle >$}\kern-0.6em
\lower0.40ex\hbox{{$\scriptstyle \sim$}}}}
\def\ls{\mathrel{\raise0.35ex\hbox{$\scriptstyle <$}\kern-0.6em
\lower0.40ex\hbox{{$\scriptstyle \sim$}}}}

\slugcomment{Accepted: 13 February 2007}

\shorttitle{An Increase in the faint red galaxy population in massive clusters since z$\sim$0.5}
\shortauthors{Stott et al.}

\begin{document}

\title{An Increase in the faint red galaxy population in massive clusters since z$\sim$0.5}
\author{J.\,P.\ Stott\altaffilmark{1}, 
Ian Smail,\altaffilmark{1}
A.\,C.\ Edge,\altaffilmark{1}
H.\ Ebeling,\altaffilmark{2}
G.\,P.\ Smith,\altaffilmark{3}
J.-P.\ Kneib\altaffilmark{4} \&
K.\,A. Pimbblet\altaffilmark{5}
}

\altaffiltext{1}{Institute for Computational Cosmology, Department of
  Physics, Durham University, South Road, Durham DH1 3LE, UK. E-mail:
  j.p.stott@durham.ac.uk}
\altaffiltext{2}{Institute for Astronomy, 2680 Woodlawn Drive, Honolulu, HI 96822}
\altaffiltext{3}{School of Physics and Astronomy, University of
        Birmingham, Edgbaston, Birmingham, B15 2TT, UK}
\altaffiltext{4}{Laboratoire d'Astrophysique de Marseille, Traverse du
        Siphon -- B.P.8 13376, Marseille Cedex 12, France}
\altaffiltext{5}{Department of Physics, University of Queensland, Brisbane, QLD 4072, Australia}
\label{firstpage}

\begin{abstract}
We compare the luminosity functions for red galaxies lying on the restframe $(U-V)$
color-magnitude sequence in a homogeneous sample of ten X-ray luminous
clusters from the MACS survey at $z\sim 0.5$ to a similarly selected
X-ray cluster sample at $z\sim 0.1$. We exploit deep {\it Hubble Space
Telescope} ACS imaging in the F555W and F814W passbands of the
central 1.2-Mpc diameter regions of the distant clusters to measure
precise colors for the galaxies in these regions and statistically
correct for contamination by field galaxies using observations of blank
fields.  We apply an identical analysis to ground-based photometry of
the $z\sim 0.1$ sample. This comparison demonstrates that the number of
faint, $M_V\sim -19$, red galaxies relative to the bright population
seen in the central regions of massive clusters has roughly doubled
over the 4\,Gyrs between $z\sim 0.5$ and $z\sim 0.1$.  We quantify this
difference by measuring the dwarf to giant ratio on the red sequence
which increases by a factor of at least $2.2\pm 0.4$ since
$z\sim0.5$. This is consistent with the idea that many faint, blue
star-forming galaxies in high density environments are transforming
onto the red sequence in the last half of the Hubble time. 
\end{abstract}

%
%
%
\section{Introduction}
\label{sec:intro}
Recent studies have precisely quantified the variation in the
photometrically classified galaxy population as a function of
environment at low redshifts (\citealt{Hogg2004}).  These studies
separate galaxies into red, passive or blue, star-forming systems and
find that the proportion of the latter decreases in higher density
regions in the local Universe \citep{Baldry2006}.  The
identification of the physical process responsible for this trend is
still contentious, in part because it is likely that a number of
processes contribute in different environments, at different epochs and
acting on galaxies of different luminosities/masses.  The presence of a range
of potential evolutionary pathways linking star forming and passive
galaxies may be reflected in the diversity of star formation histories
derived for passive early type galaxies.  While the formation of
luminous early type galaxies in clusters has been interpreted in terms
of a narrow range in star formation histories
\citep{Bower1992,AragonSal1993,vanDok1998,Stanford1998}, there is
evidence of much more variety in lower luminosity systems
\citep[see][for a review]{Ferguson1994}. Several lines of evidence
illustrate this, for example \cite{Poggianti2001} find a broad
range in ages but a slight decrease in age for fainter dwarf galaxies
in the Coma cluster. \cite{Smail2001} reached a similar conclusion
for the luminosity weighted ages of low luminosity early type galaxies
in the $z=0.18$ cluster A\,2218.  An even wider variety is found in
much lower luminosity systems, \cite{Grebel1999} finds that star
formation timescales and ages vary significantly for the local group
dwarf ellipticals.  Indeed, this same effect may underlie the varying
morphological mix seen in the passive galaxy populations in clusters,
where there is a claim of a deficit of early type disk galaxies
(predominantly S0) in distant clusters (\citealt{dres1997}). Thus it
appears that passive galaxies may be formed via a number of different
processes and that the mixed nature of the population may be most easily discerned at the lowest luminosities.

These different pathways may also result in a changing passive galaxy
population in clusters at different redshifts.  The most fundamental
measure of the transformational processes forming passive galaxies in
high density regions is to look at the build up of the luminosity
function of this population. \cite{deluc2004} therefore investigated
this scenario by measuring the color-magnitude relation in four
optically selected $z\sim0.75$ clusters and comparing these to the
nearby Coma cluster.  They find that the high redshift clusters exhibit
a deficiency in low luminosity red galaxies compared to Coma.
Similarly, \cite{kodama2004} found a deficit of red sequence galaxies when
looking at $z\sim1$ high density regions in the Subaru/{\it XMM-Newton}
Deep Survey.  This leads them to conclude that many faint red galaxies
in clusters have only moved onto the red sequence since $z\sim0.75$
(see also \citealt{delucia2007}). This result is controversial as \cite{Andreon2006} claims there is no evidence for this red sequence build up when comparing the luminosity function of a cluster at $z=0.831$ to those of local clusters.  

In this paper we aim to test these results by comparing the
evolution in the luminosity function of galaxies on the red sequence
in two well-defined samples of X-ray selected clusters at $z\sim
0.5$ and $z\sim 0.1$. We employ {\it Hubble Space Telescope (HST)}
optical imaging of ten $z\sim0.5$ X-ray luminous clusters and compare
these to a similar sample at $z\sim0.1$ to
examine evolution in the faint end of the red sequence luminosity
function. We adopt a $\Lambda$CDM cosmology ($\Omega_{M}=0.3$,
$\Omega_{Vac}=0.7$, $H_{0}=70$\,km\,s$^{-1}$\,Mpc$^{-1}$) in which the
lookback times to $z=0.13$ and 0.54 are 1.6 and 5.3\, Gyrs and the
angular scales are such that $1''$ corresponds to 2.3 and 6.4\,kpc respectively.
An AB magnitude system is used throughout this paper.

\section{Observations and Reduction}

Our analysis employs restframe optical imaging of two cluster samples
at $z\sim 0.1$ and $z\sim 0.5$. The $z\sim0.5$ sample are selected from
the MAssive Cluster Survey (MACS, \citealt{MACS2001}). MACS is a
survey of distant X-ray luminous, ($L_{X}>10^{44}$\,erg\,s$^{-1}$) and
therefore massive, galaxy clusters selected from the {\it ROSAT}
All-Sky Survey.  The ten clusters in this sample, along with a further
two for which archival {\it HST} observations exist, are the twelve
most distant clusters, $z\sim 0.5$--0.7, from MACS and all have X-ray
luminosities of $L(\rm $0.1--2.4\,KeV$)>15\times
10^{44}$\,erg\,s$^{-1}$ (Table~1).  These ten clusters were imaged with
the Advanced Camera for Surveys (ACS) Wide Field Channel (WFC) on {\it
HST} during Cycle~12 (GO\#9722).  Each cluster was imaged for two
orbits (4.5\,ks) through both the F555W (V$_{555}$) and F814W
(I$_{814}$) filters.  These data were retrieved from the STScI archive
and reduced using standard STScI software ({\sc multidrizzle} v2.7) to
provide final images with 0.05$''$ sampling and good cosmetic
properties.

We extracted the photometry from the ACS images using {\sc SExtractor}
\citep{sextractor1996} run in dual-image mode so that photometric
information from the $V_{555}$-band was extracted for all sources
detected on the $I_{814}$-band image and ensuring that we have precise
aperture photometry for even crowded sources. Throughout this paper we
use 1.4-$''$ apertures (9\,kpc diameter) to calculate the
$(V_{555}-I_{814})$ color and the `Best' magnitude is used for the
total $I_{814}$-band magnitude (these correspond closely to restframe
$(U-V)$ colors and $V$ absolute magnitudes).

The low-redshift, $z\sim0.1$, comparison sample for our analysis comes
from the Las Campanas/AAT Rich Cluster Survey (LARCS,
\citealt{LARCS2001}, 2006). This survey obtained panoramic, ground-based $B$-
and $R$-band imaging of ten X-ray luminous ($L_{X}>5\times
10^{44}$\,erg\,s$^{-1}$) clusters at $z=0.08$--$0.15$ selected from the
{\it ROSAT} All-Sky Survey (Table~\ref{tab:sample}).  The observations
and their reduction and analysis are described in detail in
\cite{LARCS2001}.  Here we use 4-$''$ $(B-R)$ colors (corresponding
to restframe $(U-V)$ colors in 9\,kpc apertures) and total magnitudes
derived from these restframe $\sim V$-band selected galaxy catalogues
in our analysis.  These data are thus well-matched to the observations
of the distant sample: the absolute $V$-band surface brightness limits
are $\mu_V = -15.4$ and $-15.8$ mag.\ arcsec$^{-2}$ for LARCS and MACS
respectively, with spatial resolution of 2.8\, and 1.0\,kpc. In our analysis we only consider the inner parts of each cluster,
within a radius of 600\,kpc of the cluster center as identified from
the X-ray emission (usually corresponding to the position of the
brightest cluster galaxy) as this provides uniform coverage across both
the LARCS and MACS datasets.  

Both the MACS and LARCS cluster samples were selected from the same
X-ray survey and the luminosities for the clusters are sufficiently
bright that they should correspond to some of the most extreme
environments at their respective epochs.  The median X-ray luminosities
of the high- and low-redshift samples are 17.0 and
$7.3\times10^{44}$\,ergs\,s$^{-1}$ respectively, corresponding to a
difference of less than a factor of two in the typical total mass
\citep{reip2002}. However, an important issue to address is that the
mass of the $z\sim0.5$ progenitors of the LARCS clusters would be even
lower than the MACS clusters. Using the results of \cite{tormen1998} we
see that the progenitors of the LARCS clusters at $z\sim0.5$ would be $\sim3.5 \times$ less massive than the MACS sample (with
corresponding X-ray luminosities of
$\sim2\times10^{44}$\,ergs\,s$^{-1}$, \citet{reip2002}).  There is no
evidence for strong variations in the galaxy luminosity function
between clusters spanning such a relatively modest difference in
typical mass \citep{deprop1999}. In addition, \cite{Wake2005} see no variation in the blue galaxy fraction in clusters covering a similar range in mass. Combining these two results, we therefore expect that any
differences between the galaxy populations in these two samples will
primarily reflect evolutionary differences between $z\sim 0.5$ and
$z\sim 0.1$.

Finally, to better define the evolutionary trends we are searching for,
we also include similar observations of seven additional clusters in
our analysis of the dwarf-giant ratio in \S\ref{sec:gdr}.  We include a
low redshift point from $U$- and $V$-band observations of the Coma
cluster from a dataset of known members \citep{Godwin1983}. This is well matched to our main MACS and
LARCS samples.  Next, we include four additional clusters which are at
redshifts intermediate between the LARCS and MACS samples and a further
two clusters at redshift beyond the MACS sample (see
Table~\ref{tab:sample}).  These clusters all have deep archival
observations with {\it HST} in the F555W/F606W and F814W filters (from GO
projects 9033, 9290, 9722, 10325, 10491, 10872 and 10875) and we obtain
the data from the {\it HST} archive. We note that the {\it HST} ACS
observations of the four intermediate clusters do not reach the 600-kpc
radius adopted for our analysis and so we have corrected the
dwarf-giant ratio using the average observed radial trend in this
ratio for the composite MACS sample. The dwarf-giant ratio is observed to increase in value from the cluster center outwards and flatten at larger radii. For the most extreme case, cluster Abell 1703 at $z=0.258$, this correction gives an enhancement of ~15\% in dwarf-giant ratio. In addition we caution that the
K-corrections for the additional {\it HST} clusters are not as
well-matched to restframe $(U-V)$ as those for the LARCS or MACS
samples and so there may be systematic uncertainties related to these
data points.
\section{Analysis and Discussion}
\label{sec:DR}
\begin{figure*}
\centering
\scalebox{0.6}[0.6]{\includegraphics*{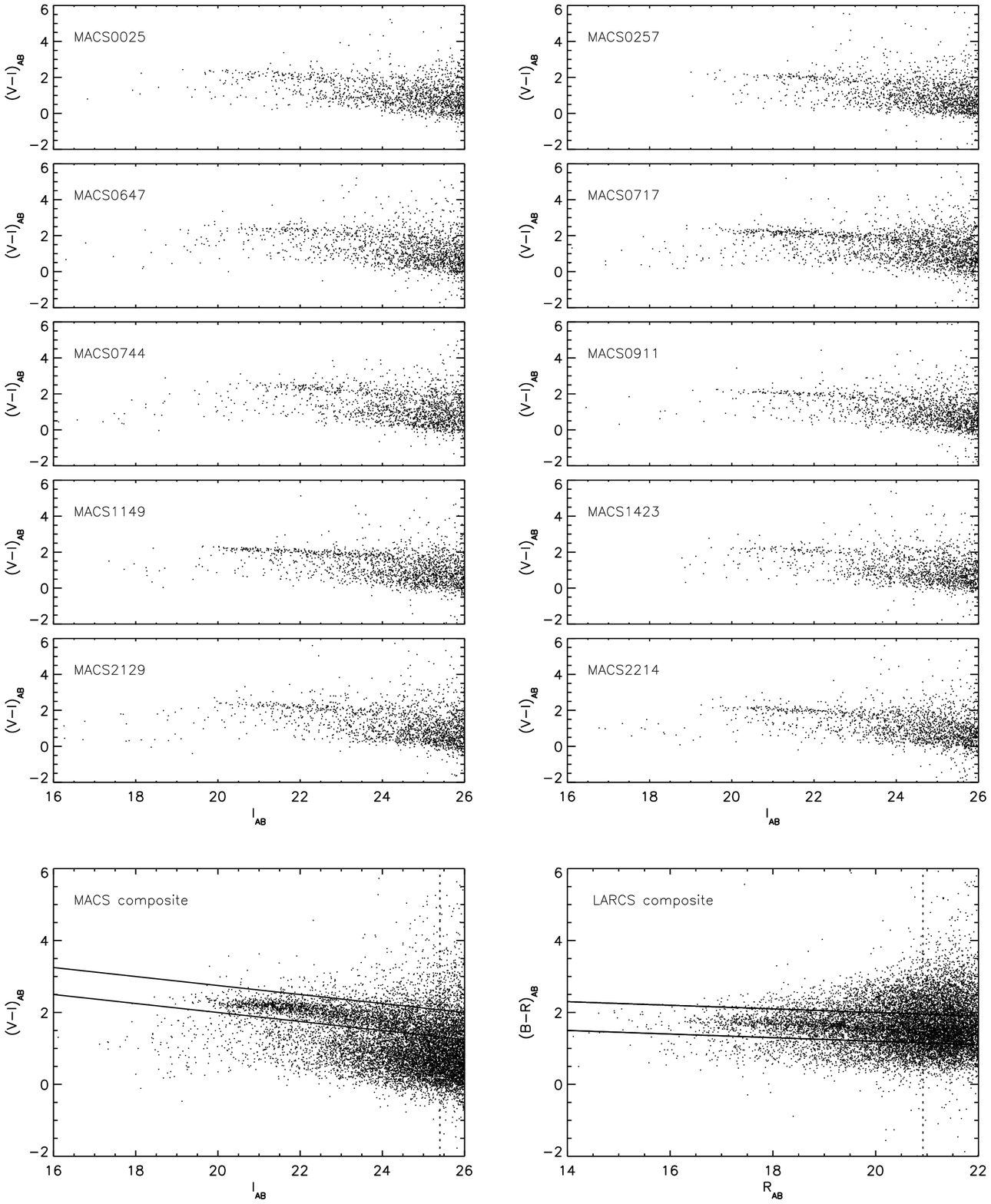}}
\caption{{\it Upper ten panels}: The individual $(V_{555}-I_{814})$ color-magnitude
diagrams for the MACS clusters. {\it Bottom left panel}: The combined
color-magnitude diagram for the MACS sample (corresponding to
restframe $(U-V)$--$V$), all clusters have been
K-corrected to $z=0.54$.  The solid lines show the limits used to
define and select the red sequence in the combined clusters and the
dotted line is the 5-$\sigma$ limit $I=25.4$.  {\it Bottom right
panel}: The combined $(B-R)$--$R$ (restframe $(U-V)$--$V$)
color-magnitude diagram for the LARCS sample. The
dotted line denotes the 5-$\sigma$ limit of $R=20.92$ and all clusters
have been K-corrected to $z=0.13$.  Again the solid lines show the limits
used to select the red sequence in the combined clusters, these are
transformed from the equivalent boundaries for the MACS sample as
described in the text. Color-magnitude plots for individual clusters in
the LARCS survey can be found in \cite{pimb2002}.}

\label{fig:Mindcmr}

\end{figure*}

We show in Fig.~\ref{fig:Mindcmr} the color-magnitude diagrams for the
individual MACS $z\sim 0.5$ clusters (similar plots for the individual
LARCS clusters are given in \citealt{pimb2002}).  These display
strong linear features in the distributions of the brighter and redder
galaxies in the fields.  These are the color-magnitude relations for
the passive, early-type cluster members \citep{Sandage1978,Bower1992}.
For the brighter galaxies in the clusters this sequence is thought to
represent variations in the metallicity and to a lesser extent age of
the stellar populations in the galaxies (\citealt{Kodama1997};
\citealt{Terl2001}).

As our MACS and LARCS cluster samples are homogeneously selected and
observed and cover a small range in redshift ($\sigma z/z \sim 0.15$),
we are able to combine them to reduce the influence of variations in
field contamination and to improve the statistics of our analysis.  We
show in the two lower panels of Fig.~\ref{fig:Mindcmr} the combined
color-magnitude relations for the MACS and LARCS samples. The combined
color-magnitude plots were created by correcting the data to the median
redshifts of the MACS and LARCS samples, $z=0.54$ and $z=0.13$
respectively, using the K-corrections from a \cite{BC2003} solar metallicity, simple stellar population, $z_{f}=5$ model and the
appropriate distance modulae.  We also define limits to the
color-magnitude relation in these combined samples to allow us to
quantitatively compare them. We define the limits on the basis of an error weighted two-parameter fit to the slope of the combined MACS red sequence with a stripe width of 0.75 mags to include the observed scatter. The corresponding boundaries
for the LARCS red sequence are then determined by K-correcting and
color converting the limits from the MACS sample using
\cite{BC2003} and the formulae given in \cite{Skiff2003} and
\cite{Natali1994}. A correction for the observed change in red
sequence slope between the two epochs is also included by using the gradient found from error weighted fitting to the combined LARCS red sequence.  We plot on
Fig.~\ref{fig:Mindcmr} the corresponding color boundaries for the two
samples.  It is these red sequences that are used below to estimate the
combined red sequence luminosity functions and the relevant field
correction.

\subsection{Field Correction}
\label{sec:FC}

The removal of field galaxies from our samples is crucial to provide a
clean measurement of the cluster luminosity function.  For the MACS
{\it HST} sample we used blank fields from The Great Observatories
Origins Deep Survey North (GOODS, \citealt{goods}).  These provide
estimates of the correction for field contamination in each bin in our
color-selected luminosity functions. The variation in this correction
between independent 16 arcmin$^2$ subregions of the 112 arcmin$^2$ blank field is included in the
luminosity function errors. The correction for field contamination for
the ground-based observations of the LARCS sample is calculated in a
similar way.  Here, however, we make use of the fact that the
panoramic, 2-degree diameter, imaging of these clusters extends into
the field population surrounding the clusters and so we can use the
outskirts of the images to determine the field contamination. These
estimates have been shown to be robust by \cite{pimb2002}. Again we
determine the reliability of our field corrections by dividing the
total 4600 arcmin$^2$ field region into 290 arcmin$^2$ subregions and determining the scatter in these independent areas. These field subregions are on scales comparable to the regions of the MACS and LARCS clusters we analyse. The typical 1-sigma errors for the field correction are in the region of 10 percent for MACS and 20 percent for LARCS. We propagate these uncertainties through our entire
analysis. 

An additional source of contamination comes from higher redshift
galaxies that have been gravitationally lensed by our clusters. This
would effectively alter the field contamination in our sample, either
increasing or decreasing it depending upon the slope of the number
counts \citep{taylor1998}.  To estimate the scale of this
effect we compare the number counts of galaxies on the color-magnitude
plane which are just redward of the cluster's color-magnitude sequence
(and hence should be at higher redshifts) to similar regions of the
color-magnitude plane for the blank fields.  This provides an estimate
of the potential enhancement in the surface density of red galaxies in
the cluster centres due to lensing of $\sim$0.1 percent.  Although only a very
small effect, we include this factor in the field correction.

\subsection{Luminosity Function}
\label{sec:LF}

The field-corrected luminosity functions for the two composite cluster
samples are shown in Fig.~\ref{fig:Nhist}.  Luminosity functions are
traditionally fitted with a single Schechter function
\citep{schec}. Recent papers on clusters, however, have instead fitted
a Gaussian to the bright end of the luminosity function and a Schechter
function to the faint end as these give a better fit to observations
\citep{Thompson1993,Dahlen2004}. We compare the single Schechter to the
Gaussian+Schechter parameterisation of the luminosity function for
galaxies on the red sequences in the LARCS and MACS samples.  To avoid
incompleteness we only consider the luminosity function down to the
K-corrected 5-$\sigma$ limit of the highest redshift cluster in the
shallower LARCS data ($R=20.92$ from A\,3888 corresponding to
$M_V=-17.75$).  Table~\ref{tab:param} contains the best-fit Schechter
and Gaussian parameters converted to absolute $V$-band magnitudes using
the method described above and the reduced $\chi^{2}$ for these fits.
The errors quoted here are estimated using a bootstrap method.

When fitting combined Gaussian+Schechter functions we fix the mean
magnitude and width of the Gaussian components (Table~\ref{tab:param}),
within their errors, so as to constrain the evolution of the bright-end
Gaussian between $z\sim0.5$ and $z\sim0.1$. We do this as the
luminosity evolution of galaxies in the bright end of the LF is
well-constrained from fundamental plane studies \citep{vanDok1998} and
so we can focus on changes in the faint-end. The passive evolution of
the luminosity between the two epochs is fixed as 0.3 magnitudes from
\cite{vandok2001}.

We find that both a Gaussian+Schechter or a Schechter function give
acceptable fits to our two samples.  Both parametric forms also
demonstrate the same difference between the two samples: an increase in
the number of faint red galaxies compared to the brighter red
population at lower redshifts.  For the Schechter fits this is shown by
the steeper faint end slope ($\alpha$) in the LARCS clusters than MACS
clusters, $-1.11\pm0.06$ versus $-0.91\pm0.02$ respectively, a
difference of approximately $3.2\sigma$. We find no evidence for
variations of the form of the luminosity function looking at either the
bluest or reddest halves of the color-magnitude sequence or between
different clusters when ranked by richness.

For the Gaussian+Schechter fits, the change in the luminosity function
is shown in part by the relative normalisation of the faint Schechter
and bright Gaussian components, $\phi^*$/Amp, which decreases from $1.94\pm0.58$ in LARCS to $1.26\pm0.67$ for the MACS sample.  However,
the covariance between the parameters in the two components functional
fits makes such a comparison complex to interpret and so we turn to
another, shape-independent, parameterisation of the relative numbers of
faint and luminous galaxies: the dwarf-giant ratio.

\subsection{Dwarf-Giant Ratio}
\label{sec:gdr}

The results from \S\ref{sec:LF} are difficult to interpret in part
because the form of the luminosity function is complex and its
evolution is uncertain. A far simpler approach to quantify the relative
evolution of the numbers of bright and faint galaxies is to use the
ratio of the number of dwarfs to giants along the red sequence: the
Dwarf-Giant Ratio (DGR) which provides a single number to parameterize
the distribution of galaxy luminosities within a population.  The
variation of this quantity (or its inverse GDR) with distance from
cluster center, density and cluster richness have been studied by a
number of workers \citep{Driver1998,Dahlen2004,Goto2005}.  Therefore to
provide a shape-independent comparison of the red galaxy populations in
the MACS and LARCS clusters we have measured the DGR.

\begin{figure*}
\centering
\scalebox{0.5}[0.5]{\includegraphics*{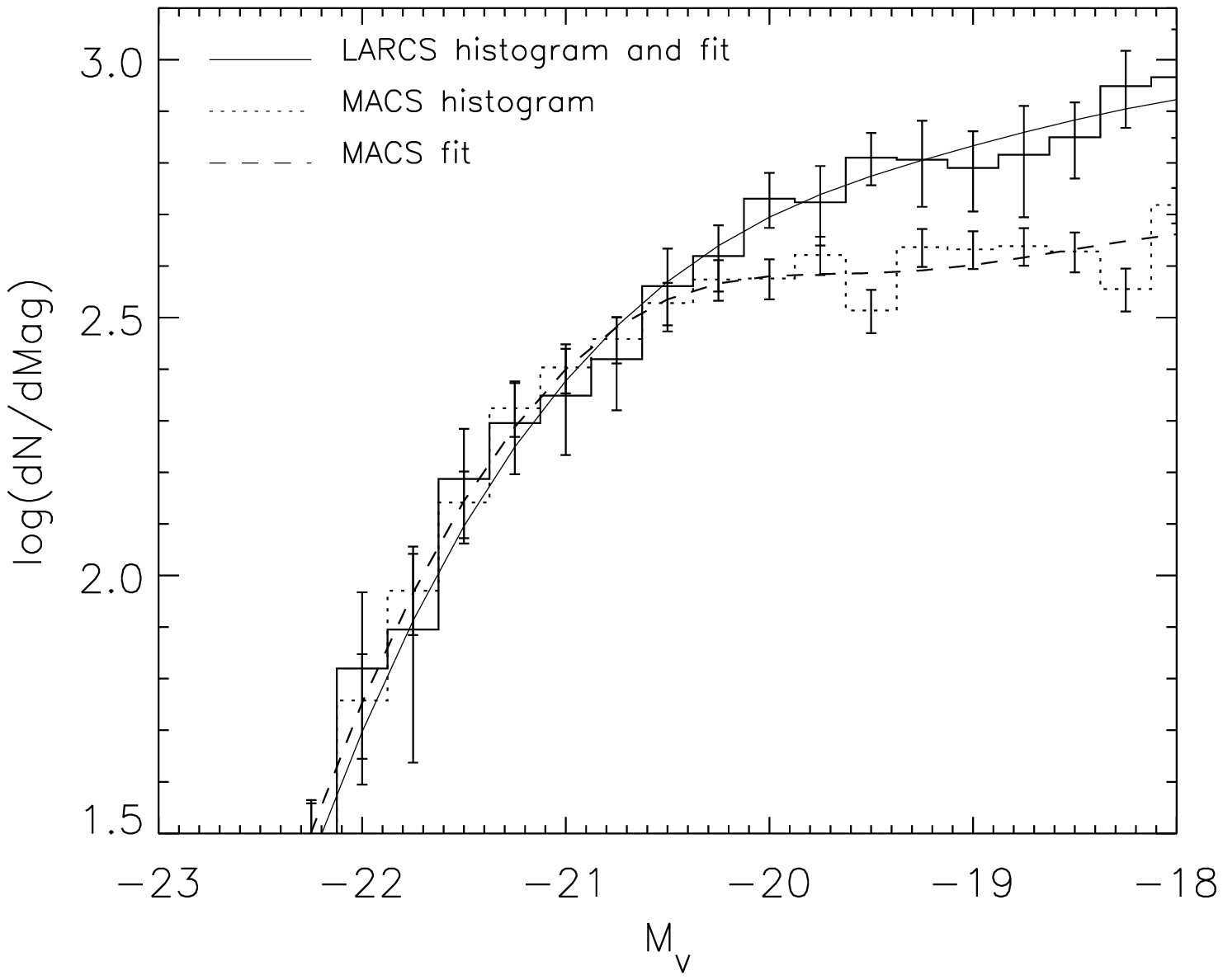}\includegraphics*{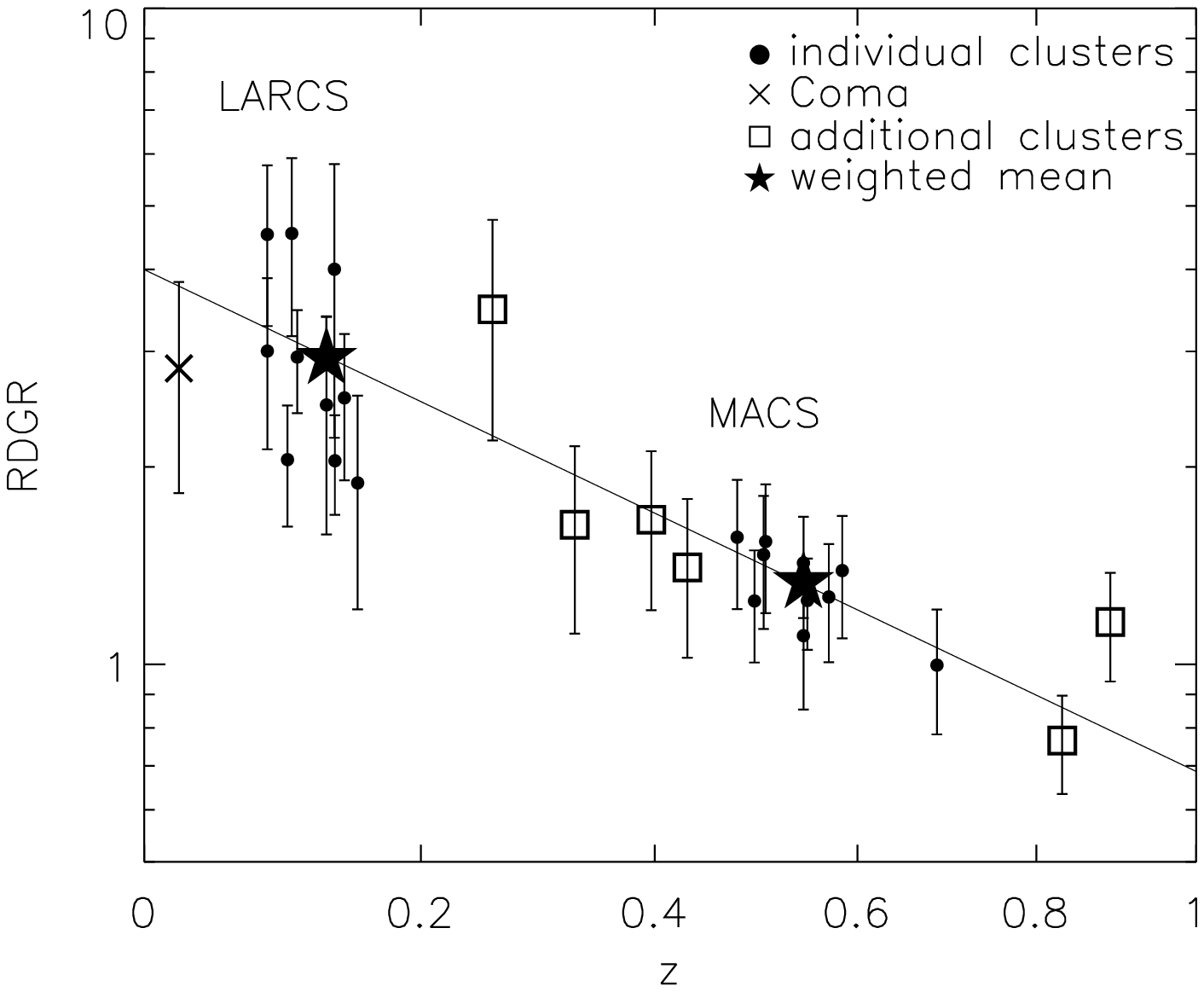}}
\caption{{\it left panel}: The luminosity functions in the restframe $V$-band for the red sequence galaxies in the combined MACS
and LARCS samples (normalized to the bright-end of the LARCS
sample). We plot the background-corrected data and our best-fit
Gaussian+Schechter function fits. The errors are a combination of the
Poisson uncertainty and the field correction error.  Note the excess of
faint red galaxies in the lower redshift LARCS sample, compared to the more
distant MACS sample. {\it right panel}: The variation in the red
sequence Dwarf-Giant Ratio (RDGR) with redshift for clusters in our two
samples within 600\,kpc of the cluster centre and brighter than
$M_{V}=-17.75$. We also plot the weighted mean values for
each of the MACS and LARCS sample.  A fit of the form $(1+z)^{-\beta}$
to the MACS and LARCS points is plotted and yields $\beta=2.5\pm0.5$.
For comparison we show the equivalent measures for six additional high
redshift clusters and Coma (RDGR$=2.8\pm1$), which follow the same trend.}
\label{fig:Nhist}
\end{figure*}

The boundary between giants and dwarfs is arbitrary and is usually
defined as the magnitude where the faint-end Schechter function begins
to dominate over the bright-end Gaussian \citep{Goto2005}.  Looking at
the distributions in Fig.\ref{fig:Nhist}, we therefore choose an
absolute magnitude of $M_V\sim -19.9$ at $z=0.13$ as our dividing
point. This absolute magnitude brightens to $M_V\sim -20.2$ at
$z=0.54$ as we take into account the passive evolution models of
\cite{vandok2001} (we have confirmed that our results are not sensitive
to applying this factor).  To ensure that our measurements of the DGR
in the different samples are not effected by incompleteness we adopt
the same faint-end limit as for fitting the luminosity functions
($M_V\sim -17.75$) and as in \S\ref{sec:LF} we only consider galaxies
within a 600\,kpc radius of the center of each cluster. The limits chosen for our DGR analysis are comparable to those of \cite{deluc2004}.

Fig.~\ref{fig:Nhist} shows the variation of the DGR on the red sequence
(RDGR) with redshift.  The RDGR increases with cosmic time and we
attribute this to an increasing number of dwarfs on the red sequence.
We determine weighted mean RDGRs for the MACS sample of $1.33\pm 0.06$
and $2.93\pm 0.45$ for LARCS, a difference of $3.7\sigma$. This
corresponds to an increase in RDGR of a factor of $2.2\pm 0.4$ from
$z=0.54$ to $z=0.13$, or a lookback interval of 4\,Gyrs.  An
alternative way to look at this evolution is the variation in the
proportion of integrated red light at the faint-end of the
color-magnitude relation, which increases by a factor of $1.46\pm0.14$
from $z=0.54$ to $z=0.13$.  This means that the stellar mass in the
passive dwarf population ($M_V\lesssim -20$) now almost equals that in
luminous cluster galaxies, whereas at $z\sim 0.5$ the giants dominated
the total $V$-band luminosity from galaxies on the color-magnitude
relation. The errors shown are a combination of the Poisson uncertainty
and the field correction error.

We note that the LARCS RDGR errors are larger than those for MACS mainly
due to the fact that at $z\sim0.1$ the field galaxies and the faint end
of the red sequence occupy the same region of $(B-R)$--$R$
color-magnitude space increasing the field correction error at the
faint end (see Fig. \ref{fig:Mindcmr}). In contrast the faint red
sequence in the MACS sample at $z\sim0.5$ is much easier to distinguish
from the field using $V$ and $I$ bands.

To parameterise the evolution in the RDGR we fit a $(1+z)^{-\beta}$
power-law to the LARCS and MACS samples in Fig.~\ref{fig:Nhist}.  We
see that this provides a good description of the evolution for
$\beta=2.5\pm0.5$, with all of the clusters consistent with the
fit. This confirms that the luminosity function of the red sequence in
the central regions of massive clusters appears uniform with no clear
evidence from our study of real cluster-to-cluster variations although the errors on individual clusters are large.  We also
find that the six additional clusters and Coma (RDGR$=2.8\pm1$) follow the same trend defined
by the MACS and LARCS samples. \cite{deluc2004}, who use a similar definition of DGR, found a value of 1.23 for clusters at $z=0.75$ and a value of 2.9 for Coma which are in good agreement with the trend we observe. The trend is also in qualitative agreement with the work of \cite{kodama2004}.

\section{Conclusions}

Our analysis of the red galaxy populations in X-ray luminous clusters
shows clear differences in the form of the luminosity function over the
redshift range $z=0.1$--0.5. These changes reflect an increase in the
proportion of dwarf to giant galaxies in the population since
$z\sim0.5$ which we attribute to an increase in the number of dwarfs on
the red sequence.  We quantify this evolution using the shape
independent estimate of the relative evolution of the faint end of the
luminosity function, the red sequence Dwarf-Giant ratio (RDGR), which
shows an increase by a factor of $2.2\pm 0.4$ between $z=0.54$ and
$z=0.13$.  This is equivalent to an increase of $1.46\pm 0.14$ in the
relative $V$-band luminosity (or stellar mass) in faint red galaxies
with $M_V\lesssim -20$ compared to brighter systems over this period.
This increase means that in local clusters, the luminosity contributed
by giant and dwarf galaxies is comparable, whereas at $z=0.5$ the
giants were the dominant population on the color-magnitude relation. 

Our results show that there is significant evolution since $z\sim 0.5$
in the faint passive galaxy population in a well-defined sample of
X-ray luminous clusters.  This agrees with the early results from
\cite{deluc2004} and \cite{kodama2004} on red galaxies in a more
diverse range of structures. However, this is in disagreement with the work of \cite{Andreon2006} who sees no such evolution. This disagreement may be simply due to Andreon's use of a single cluster, as our analysis shows a large cluster to cluster scatter but with large errors (due to an uncertain field correction). We conclude that a large proportion of
the passive galaxy population at the faint end of the color-magnitude
sequence in local clusters either did not reside in similar,
high-density environments 5-Gyrs ago (at $z\sim 0.5$) or if they were
present in these regions then they had significantly bluer colors
(suggesting they were actively star forming) and so do not fall within
the color-magnitude relation.

Clusters in the mass range studied in this work are expected to have
roughly doubled their masses since $z\sim 0.5$ (Tormen 1998) and hence
many of the faint red galaxies (or at least their progenitors) may have
arrived in the core regions of the cluster between $z\sim 0.5$ and
$z\sim 0.1$.  However, we believe that the major driver of the
evolution we see is the transformation of blue, star-forming galaxies
into passive, red systems which lie on the color-magnitude relation.
If correct, this suggests that there will be an increasing diversity in
the star formation histories of passive galaxies at $M_V\lesssim -20$ in
intermediate redshift clusters (at $z\sim 0.3$--0.4), and studies of
age-sensitive indicators at these depths may uncover evidence for
recent star-formation activity within these galaxies (e.g. \citealt{Smail2001}).

We end by noting that studies such as this one can be extended to a wider
range of environments and redshifts out to $z\sim1$ using the data
from the UKIRT Infrared Deep Sky Survey (UKIDSS, \citealt{Law2007})
in concert with the {\it XMM Newton} Large Scale Structure Survey
(XMM-LSS).  Such studies will allow the evolution  of the passive
population on the color-magnitude relation and its build-up to be
tracked as a function of environment and epoch and will demonstrate
the importance of including environmental effects when
modelling the color-magnitude relation in galaxy evolution models.

\section*{Acknowledgements}

We thank the referee for their useful comments which have improved the clarity of this paper. We also thank Michael Balogh, Tadayuki Kodama, Bianca Poggianti and David Wake for
useful discussions.  JPS acknowledges support through a Particle
Physics and Astronomy Research Council Studentship. IRS and GPS
acknowledge support from the Royal Society. HE is grateful for support from STScI through grants HST-GO-09722,  
HST-GO-10491, and HST-GO-10875.


\clearpage

%

\input{tab1.tex}
%
%
\clearpage
\input{tab2.tex}

%
%
\pagebreak

%
%
%

\end{document}

%% file: tab1.tex
\begin{table*}
\begin{center}
\caption{Details of the cluster samples used in our analysis.}
\label{tab:sample}
\begin{tabular}{lcccccc}
\hline
Cluster & R.A.\ & Dec.\ & $z$ & $L_X$~~~& N$_{Red}^1$ & DGR\\
&\multicolumn{2}{c}{(J2000)}&&($10^{44}$erg\,s$^{-1}$)&&\\
\hline
\hline
\noalign{\smallskip}
\multispan{2}{MACS $z\sim 0.5$ Sample \hfil}\\
\noalign{\smallskip}
MACS\,J0025.4$-$1222 &00 25 15.84 &$-$12 19 44 & 0.478 &12.4 &168$\pm$13 &1.56$\pm$0.35\\
MACS\,J0257.6$-$2209 &02 57 07.96 &$-$23 26 08 & 0.504 &15.4 &157$\pm$13 &1.47$\pm$0.34\\
MACS\,J0647.7$+$7015 &06 47 51.45 &$+$70 15 04 & 0.584 &21.7 &183$\pm$14 &1.39$\pm$0.29\\
MACS\,J0717.5$+$3745 &07 17 31.83 &$+$37 45 05 & 0.548 &27.4 &321$\pm$18 &1.25$\pm$0.20\\
MACS\,J0744.8$+$3927 &07 44 51.98 &$+$39 27 35 & 0.686 &25.9 &173$\pm$13 &1.00$\pm$0.22\\
MACS\,J0911.2$+$1746&09 11 10.23 &$+$17 46 38 & 0.506 &13.2 &169$\pm$13 &1.54$\pm$0.34\\
MACS\,J1149.5$+$2223 &11 49 34.81 &$+$22 24 13 & 0.544 &17.3 &266$\pm$16 &1.43$\pm$0.25\\
MACS\,J1423.8$+$2404 &14 23 47.95 &$+$24 04 59 & 0.544 &15.0 &155$\pm$13 &1.11$\pm$0.25\\
MACS\,J2129.4$-$0741 &21 29 25.38 &$-$07 41 26 & 0.570 &16.4 &194$\pm$14 &1.27$\pm$0.26\\
MACS\,J2214.9$-$1359 &22 14 56.51 &$-$14 00 17 & 0.495 &17.0 &215$\pm$15 &1.25$\pm$0.24\\
\noalign{\medskip}
\noalign{\smallskip}
\multispan{2}{LARCS $z\sim 0.1$ Sample \hfil}\\
\noalign{\smallskip}
Abell\,22&00 20 38.64&$-$25 43 19& 0.142&5.3 & 220$\pm$21 & 2.55$\pm$0.64\\
Abell\,550&05 52 51.84&$-$21 03 54& 0.099&7.1 & 277$\pm$22 & 2.05$\pm$0.43\\ 
Abell\,1084&10 44 30.72&$-$07 05 02& 0.132&7.4 & 111$\pm$18 & 4.00$\pm$1.79\\
Abell\,1285&11 30 20.64&$-$14 34 30& 0.106&5.45& 391$\pm$25 & 2.94$\pm$0.53\\
Abell\,1437&12 00 25.44&$+$03 21 04& 0.134&7.7 & 376$\pm$25 & 2.04$\pm$0.35\\
Abell\,1650&12 58 41.76&$-$01 45 22& 0.084&7.8 & 182$\pm$20 & 3.00$\pm$0.88\\
Abell\,1651&12 59 24.00&$-$04 11 20& 0.085&8.3 & 232$\pm$21 & 4.52$\pm$1.24\\
Abell\,1664&13 03 44.16&$-$24 15 22& 0.128&5.34& 127$\pm$18 & 2.49$\pm$0.91\\
Abell\,2055&15 18 41.28&$+$06 12 40& 0.102&4.8 & 201$\pm$21 & 4.54$\pm$1.37\\
Abell\,3888&22 34 32.88&$-$37 43 59& 0.153&14.5& 124$\pm$19 & 1.89$\pm$0.68\\
\noalign{\medskip}
\multispan{2}{Additional Clusters \hfil}\\
\noalign{\smallskip}
Cl\,J0152$-$1357     & 01 52 43.91 &$-$13 57 21  & 0.831 &5.0   &276$\pm$17           &0.77$\pm$0.14\\
MACS\,J0451.9$+$0006 & 04 51 54.63 &$+$00 06 18  & 0.430 &10.4  &179$^{\star}\pm$14   &1.41$\pm$0.30\\
MACS\,J0712.3$+$5931 & 07 12 20.45 &$+$59 32 20  & 0.328 &6.8   &90$^{\star}\pm$10    &1.63$\pm$0.44\\
Cl\,J1226.9$+$3332   & 12 26 58.13 &$+$33 32 49  & 0.890 &20.0  &232$\pm$15           &1.16$\pm$0.22\\
Abell 1703           & 13 15 00.70 &$+$51 49 10  & 0.258 &8.7   &94$^{\dagger}\pm$10  &3.48$\pm$1.13\\
MACS\,J1354.6$+$7715 & 13 54 19.71 &$+$77 15 26  & 0.397 &8.2   &156$^{\dagger}\pm$13 &1.66$\pm$0.37\\
\hline
\end{tabular}
\end{center}
\footnotesize{1) N$_{red}$ is the number of galaxies on the red
  sequence down to M$_{V}=-17.75$ within a 600\,kpc radius of cluster
  centre except: $\star$ within a 400\,kpc radius, $\dagger$ within a
  300\,kpc, due to the size of the ACS image. The limited radius of
  MACS\,1354 is due to size of ACS image and flaring on the image from a bright
  star. The LARCS redshifts are from \cite{LARCS2006}. MACS redshifts come from
  \cite{Ebeling2007}.}
\end{table*}

%% file: tab2.tex
\begin{table*}
\begin{center}
\caption{The best-fitting parameters for the luminosity function of
red sequence galaxies for the MACS and LARCS clusters.}
\scalebox{0.95}[0.95]{\begin{tabular}{lccccccc}
\hline
&$\chi^{2}/\nu$&\multicolumn{3}{c}{Schechter}&\multicolumn{3}{c}{Gaussian$^{1}$}\\
&&$\alpha$&$M_{V}^{*}$&$\phi^{*}$&$\langle M_{V} \rangle$&$\sigma_M$&Amp. \\
\hline
\hline
LARCS&0.28&$-1.11\pm0.05$&$-21.10\pm0.11$&$177.7\pm22.2$&...&...&...\\
&0.38&$-0.92\pm0.14$&$-19.43\pm0.17$&$331.6\pm57.2$&$-20.39\pm0.12$&$0.89\pm0.05$&$170.5\pm21.8$\\
\noalign{\smallskip}
MACS&1.29&$-0.91\pm0.02$&$-21.39\pm0.05$&$215.1\pm10.9$&...&...&...\\
&1.14&$-0.94\pm0.19$&$-20.08\pm0.67$&$223.0\pm75.4$&$-20.86\pm0.16$&$0.83\pm0.08$&$177.2\pm33.9$\\
\hline
\end{tabular}}
\end{center}
\footnotesize{1) Where the Gaussian fit parameters are not
given this is a purely Schechter function fit to the luminosity
function.}
\label{tab:param}
\end{table*}